\documentclass[a4paper,12pt]{article}
\usepackage{standalone}

\usepackage[utf8]{inputenc}
\usepackage[english]{babel}
\usepackage{amssymb}
\usepackage{amsmath}
\usepackage{mathrsfs}
\usepackage{cite}
\usepackage{url}
\usepackage[ruled]{algorithm2e}
\SetArgSty{textup}
\SetKw{Error}{error}
\SetKw{And}{and}

\usepackage{graphicx,xcolor}
\usepackage[noshell]{gnuplottex}
\usepackage[pdf]{graphviz}

\usepackage{float}


\newcommand{\Mod}{\ \mathrm{mod}\ }

\newcommand{\PriceBetter}{\vartriangleleft}

\title{\textit{glass}: ordered set data structure for client-side order books}
\author{Viktor Krapivensky\\ \textit{AKB System}}
\date{%
    April 20, 2025
}

\begin{document}

\begingroup\catcode`\#=12
\newcommand\MyHash{} 
\gdef\MyHash{#}
\endgroup

\newcommand\MyColor[1]{\MyHash#1}

\maketitle

\begin{abstract}
    The ``ordered set'' abstract data type with operations
    $\underline{\mathrm{insert}},$
    $\underline{\mathrm{erase}},$
    $\underline{\mathrm{find}},$
    $\underline{\mathrm{min}},$
    $\underline{\mathrm{max}},$
    $\underline{\mathrm{next}}$ and
    $\underline{\mathrm{prev}}$ is ubiquitous in computer science.
    It is usually implemented with red-black trees, $B$-trees, or $B^+$-trees.
    We present our implementation of ordered set based on a trie.
    It only supports integer keys (as opposed to keys of any strict weakly ordered type)
    and is optimized for market data, namely for what we call \textit{sequential locality}.
    The following is the list of what we believe to be novelties:
    \begin{itemize}
        \item \textit{Cached path} to exploit \textit{sequential locality}, and fast truncation thereof on $\underline{\textrm{erase}}$ operation;
        \item A hash table (or, rather, a \textit{cache table}) with hard $O(1)$ time guarantees on any operation to speed up key lookup (up to a pre-leaf node);
        \item Hardware-accelerated ``find next/previous set bit'' operations with \textbf{BMI2} instruction set extension on x86-64;
        \item Order book-specific features: the \textit{preemption principle} and the \textit{tree restructure} operation that prevent the tree from consuming too much memory.
    \end{itemize}
    We achieve the following speedups over C++'s standard \texttt{std::map} container:
    \textbf{6x---20x} on modifying operations,
    \textbf{30x} on lookup operations,
    \textbf{9x---15x} on real market data,
    and a more modest \textbf{2x---3x} speedup on iteration.
    In this paper, we discuss our implementation.
\end{abstract}

\section{Notation}

By $\mathbb{N}$ we mean $\{ 0, 1, 2, \ldots \}.$

By $a \Mod b,$ where $a \in \mathbb{Z}, b \in \mathbb{N}, b \ne 0,$ we mean the integer $r$ such that $0 \le r < b$ and $r \equiv a \pmod{b}.$

By $\mathbb{Z}_m$ we mean the ring of integers modulo $m.$
We identify elements of $\mathbb{Z}_m$ with elements of $\mathbb{N}$ as follows:
\begin{itemize}
    \item we identify $x \in \mathbb{Z}_m$ with the minimal $n \in \mathbb{N}$ such that
        $\underbrace{x = \hat{1} + \hat{1} + \cdots + \hat{1}}_{n \, \mathrm{times}};$
    \item we identify $n \in \mathbb{N}$ with the unique $x \in \mathbb{Z}_m$ such that
        $\underbrace{x = \hat{1} + \hat{1} + \cdots + \hat{1}}_{n \, \mathrm{times}}.$
\end{itemize}
Above, $\hat{1}$ means the multiplicative identity of $\mathbb{Z}_m.$

\section{Introduction}

The ``ordered set'' is an abstract data type which maintains a set $\xi$ of elements, which is initially empty, and has at least the following operations implemented:
\begin{itemize}
    \item $\underline{\mathrm{insert}}(\langle k, v \rangle)$: assigns $\xi \longleftarrow \xi \cup \{ \langle k, v \rangle \}$ if $\nexists \langle k', v' \rangle \in \xi : k' = k;$
    \item $\underline{\mathrm{erase}}(\langle k, v \rangle)$: assigns $\xi \longleftarrow \xi \setminus \{ \langle k, v \rangle \};$
    \item $\underline{\mathrm{find}}(k)$: if $\exists \, \langle k', v \rangle \in \xi : k' = k,$ returns $v;$ otherwise, returns the special blank symbol ``$\#$'';
    \item $\underline{\mathrm{min}}()$ and $\underline{\mathrm{max}}()$: return
    $\widetilde{\min} \, \{ k \, | \, \langle k, v \rangle \in \xi \}$
    and
    $\widetilde{\max} \, \{ k \, | \, \langle k, v \rangle \in \xi \},$ correspondingly;
    \item $\underline{\mathrm{next}}(k)$: returns $\widetilde{\min} \, \{ k' \, | \, \langle k', v \rangle \in \xi, k' > k \};$
    \item $\underline{\mathrm{prev}}(k)$: returns $\widetilde{\max} \, \{ k' \, | \, \langle k', v \rangle \in \xi, k' < k \}.$
\end{itemize}
In the list above, $\widetilde{\min}$ and $\widetilde{\max}$ specify regular $\min$ and $\max$ set operations except that they return the special blank symbol ``$\#$'' if the argument is empty.

Typically all operations are $O(\log n).$ Additionally, $\underline{\mathrm{min}}$ and $\underline{\mathrm{min}}$
may be cached and thus execute in $O(1).$

This abstract data type is ubiquitous in computer science. Examples of where it is used include databases, file systems,
and schedulers and epoll file descriptors in the Linux kernel~\cite{linux_rb}.

The C++ programming language has standard containers \texttt{std::set} and \texttt{std::map}, which are typically implemented via red-black trees~\cite{pl_cxx}.
The Java programming language provides \texttt{TreeSet} and \texttt{TreeMap} collections (under \texttt{java.utils} package), which are also implemented via red-black trees~\cite{pl_java_treemap}~\cite{pl_java_treeset}.
The Rust programming language provides \texttt{BTreeSet} and \texttt{BTreeMap} collections (under \texttt{std::collections} module)~\cite{pl_rust_set}~\cite{pl_rust_map}, which are based on $B$-trees.

There are alternatives to red-black trees and variations of $B$-trees for implementation of this abstract data type, e.g. AVL-trees and trees exploiting the structure of the keys (for example, tries and radix trees for integers and strings, van Emde Boas and fusion trees for integers).

\section{Ordered set applied to client-side order book management}

For client-side order book management, we maintain an ordered set with keys being prices and values being non-zero amounts. An pair of price and amount is called a \textit{price level}. We need to be able to handle the following queries:
\begin{itemize}
    \item $\underline{\mathrm{adjust}}(\pi, \Delta), \Delta \ne 0$: add $\Delta$ to the previous amount at price $\pi$ (if there is no previous amount at this price, set the new amount to $\Delta$). If the new amount is zero, delete this price level. $\Delta$ can be negative, but it is guaranteed that no price level has negative amount;
    \item $\underline{\mathrm{min}}()$ and $\underline{\mathrm{max}}()$: get minimum or maximum price of a level (or the special blank symbol ``$\#$'' if the order book is empty);
    \item $\underline{\mathrm{next}}(\pi)$ and $\underline{\mathrm{prev}}(\pi)$: get next or previous (after or before $\pi$) price of a level (or the special blank symbol ``$\#$'' if there is none).
\end{itemize}

$\underline{\mathrm{min}}(), \underline{\mathrm{max}}(), \underline{\mathrm{next}}(\pi)$ and $\underline{\mathrm{prev}}(\pi)$ can trivially be expressed in terms of same-name ordered set operations.
The $\underline{\mathrm{adjust}}(\pi, \Delta)$ operation can be expressed in terms of ordered set abstract data type as follows:

\begin{procedure}[H]
\DontPrintSemicolon
\caption{adjust($\pi, \Delta$)}
\KwIn{
    Price $\pi,$ signed amount $\Delta$ such that $\Delta \ne 0.$
}
\Begin{
    $A \longleftarrow \underline{\mathrm{find}}(\pi)$\;
    \uIf{$A = \#$}{
        $a \longleftarrow 0$\;
    }\Else{
        $a \longleftarrow A$\;
        $\underline{\mathrm{erase}}(\langle \pi, a \rangle)$\;
    }
    $a' \longleftarrow a + \Delta$\;
    \If{$a' \ne 0$}{
        $\underline{\mathrm{insert}}(\langle \pi, a' \rangle)$\;
    }
}
\end{procedure}

\subsection{Sequential locality and edge locality in market data}

We define \textit{sequential locality} as the closeness of the price of an event to the price of previous event;
In order words, the smaller $|\pi_i - \pi_{i-1}|,$ the greater is the sequential locality is.
We also define \textit{edge locality} as the closeness of the price of an event to the best price;
In order words, the smaller $|\pi_i - \pi_{\mathrm{best}}|,$ the greater the edge locality is.

We recorded market data on MOEX (Moscow Exchange) for instrument \textbf{CRM5} (futures contract for CNY/RUB) during the main trading session of May 20, 2025.
We then processed the recorded market data to visualize both sequential locality and edge locality:

\begin{figure}[H]
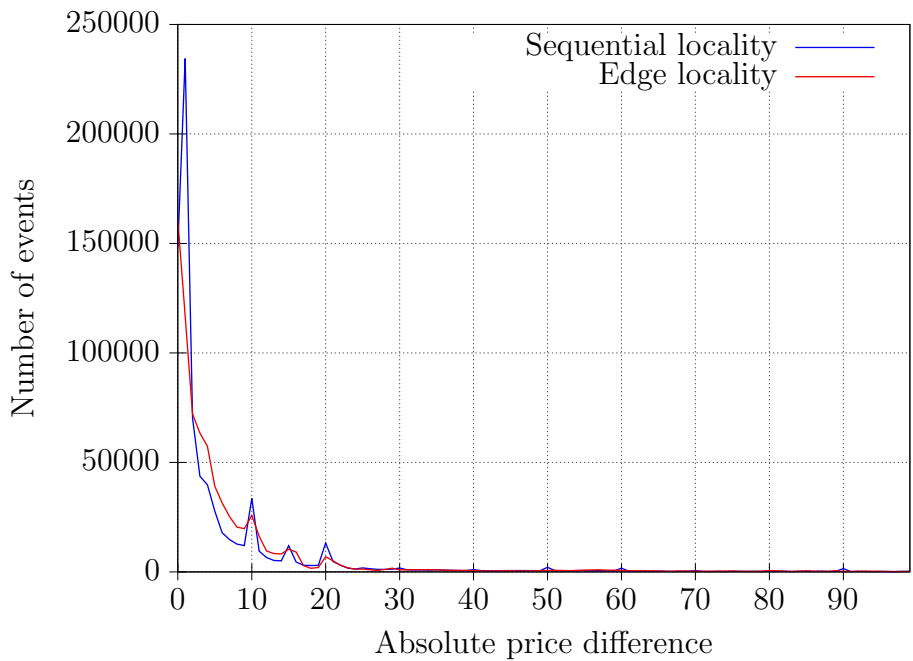

    \caption{Sequential locality and edge locality in market data}
    \centering
    \begin{gnuplot}[terminal=epslatex, terminaloptions=color]

        set style line 1 lt 1 lw 2 lc rgb '#0000ee' pt -1
        set style line 2 lt 1 lw 2 lc rgb '#ee0000' pt -1

        set xtics axis
        set ytics axis

        set grid xtics lc rgb '#555555' lw 1 lt 0
        set grid ytics lc rgb '#555555' lw 1 lt 0

        set xlabel "Absolute price difference"
        set ylabel "Number of events"

        set autoscale xfix

        plot \
            'moex-seq.txt' using 1:2 with lines ls 1 ti "Sequential locality", \
            'moex-edge.txt' using 1:2 with lines ls 2 ti "Edge locality"

    \end{gnuplot}
\end{figure}

We see that market data exhibits both strong sequential locality and strong edge locality.
Note also the peaks at what humans percept as ``nice round numbers'', e.g. 10, 15, 20, 50.

What this means is that we can cache the path to the previously accessed key to exploit sequential locality.
We could cache the path to the ``best'' (minimal or maximal, depending on the side of the order book) key to exploit edge locality,
but this would be slower because, after deletion of the best key there is no quick way to locate the next best one.
Caching both would also result in suboptimal performance because we would need to maintain two cached paths instead of one.

\section{Baseline implementation}

\subsection{The trie}

Our implementation is based on an uncompressed trie.
We fix $K,$ the number of bits in a key.
We also fix $N,$ the maximum number of children of a node. $N$ must be a power of two. We define $C = \log_2 N.$
If $C$ does not divide $K,$ we pretend the higher bits of the key are zero.
The tree, unless empty, has the unique root node.

\begin{figure}[H]
\caption{A trie}
\centering
\digraph[scale=0.6]{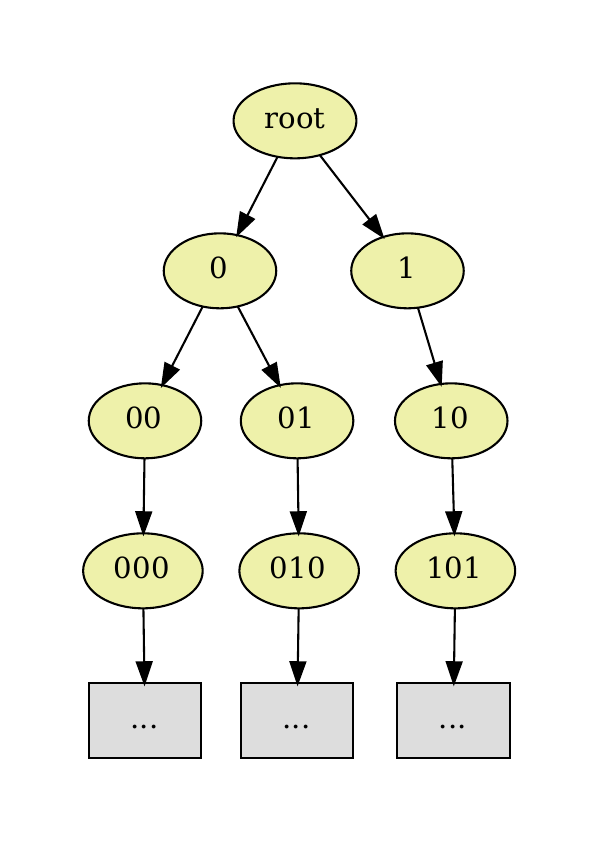}{
graph [ fontname = "DejaVu Math TeX Gyre"; ]
node [ fontname = "DejaVu Math TeX Gyre"; ]
edge [ fontname = "DejaVu Math TeX Gyre"; ]
layout = "dot";
node [
fillcolor = "\MyColor{eef1aa}"; style = "filled";
]
edge[weight=0]
"start" [ label = "root";  ]
"n0" [ label = "0";  ]
"n00" [ label = "00";  ]
"n000" [ label = "000";  ]
"n0000" [ label = "..."; fillcolor = "\MyColor{dddddd}"; style = "filled"; shape = "box"; ]
"n01" [ label = "01";  ]
"n010" [ label = "010";  ]
"n0100" [ label = "..."; fillcolor = "\MyColor{dddddd}"; style = "filled"; shape = "box"; ]
"n1" [ label = "1";  ]
"n10" [ label = "10";  ]
"n101" [ label = "101";  ]
"n1010" [ label = "..."; fillcolor = "\MyColor{dddddd}"; style = "filled"; shape = "box"; ]
start -> n0;
n0 -> n00;
n00 -> n000;
n000 -> n0000;
n0 -> n01;
n01 -> n010;
n010 -> n0100;
start -> n1;
n1 -> n10;
n10 -> n101;
n101 -> n1010;
subgraph cluster_cached_path {
label = "Cached path";
color = "\MyColor{beff8e}"; style = "filled";
}
edge[style=invis,weight=1]
;
}
\end{figure}

Above, $K = 4, \, C = 1, \, N = 2.$
We call the grey ``...'' nodes ``post-leafs'', and nodes one level higher (with three-digit labels) ``pre-leafs''.

Any node contains a $N$-bit mask indicating which children are present; array of $N$ index-pointers to children nodes, and an index-pointer to the parent node (with a special value for ``no parent'' reserved for the root node).
An iterator contains the index of a pre-leaf node and the key, the $C$ least significant bits of which specify the index of a child of the post-leaf node.

We define $\mathcal{L} = \lceil K / C \rceil,$ the distance from the root of the tree to a post-leaf.

\subsection{Index-pointers, slot allocator and multiple-node allocation}

Each index-pointer is an index into the array of nodes that a \textit{glass} maintains. Throughout this section, this array is referred to as $\mathcal{N}.$

The main reason for the introduction of index-pointers is cache locality:
if the trie can never grow to $2^{32}$ (sans two special values for ``end/not found'' and ``bad'') nodes,
it is beneficial to only store 32-bit indices (even considering that this introduces another level of indirection).
The same applies to 16-bit indices.

The data structure keeps track of free nodes via the \textit{slot allocator} principle:
\begin{itemize}
    \item free nodes are organized as a singly-linked list: a \textit{glass} maintains the ``first free node'' index-pointer,
    and one of the fields in a free node is used as the ``next free node'' index-pointer;

    \item allocating a node consists of consulting the ``first free node'' index-pointer; say its value is $p.$ If $p$ is the special ``invalid'' value, grow the nodes array and load $p$ again.
      Read the ``next free node'' field of $\mathcal{N}[p]$, call that value $q.$ Assign $q$ to the ``first free node'' index-pointer, return $p$ as the index of the allocated node.
      Note that, unless the array is grown, this is an O(1) operation.
      Note also that array growth becomes exponentially less likely as the size of a \textit{glass} grows; if full pre-allocation is used, the array never grows at all;

    \item de-allocating a node with index $p$ consists of writing the value of the current ``first free node'' index-pointer into the ``next free node'' field of $\mathcal{N}[p],$
    then replacing the current ``first free node'' index-pointer by $p.$ Note this is an O(1) operation.
\end{itemize}

See the listings below for \underline{allocate-node} and \underline{deallocate-node} functions (``$\#$'' denotes the special ``invalid'' value of an index-pointer).

\begin{procedure}[H]
\DontPrintSemicolon
\caption{allocate-node()}
\KwData{
    $\mathcal{N},$ the array of nodes.
    $\mathfrak{f},$ the ``first-free-node'' index-pointer.
}
\KwOut{
    The index-pointer to the allocated node.
}
\Begin{
    \If{$\mathfrak{f} = \#$}{
        \underline{grow-array}$()$\;
    }
    $p \longleftarrow \mathfrak{f}$\;
    $\mathfrak{f} \longleftarrow \mathcal{N}[p].\mathrm{next\_free\_node}$\;
    \Return $p$\;
}
\end{procedure}

\begin{procedure}[H]
\DontPrintSemicolon
\caption{deallocate-node($p$)}
\KwData{
    $\mathcal{N},$ the array of nodes.
    $\mathfrak{f},$ the ``first-free-node'' index-pointer.
}
\KwIn{
    $p,$ the index-pointer to the node to de-allocate.
}
\Begin{
    $\mathcal{N}[p].\mathrm{next\_free\_node} \longleftarrow \mathfrak{f}$\;
    $\mathfrak{f} \longleftarrow p$\;
}
\end{procedure}

The main drawback of such a scheme is that we never ``give back'' memory that once has been used but currently is not, to the system.
But this is widely considered a good trade-off: many user-space memory allocators, such as the one found in glibc~\cite{mem_glibc},
do not give back memory either.
Memory management in some dynamic programming languages,
such as Lua~\cite{mem_lua} (PUC Lua, i.e. the official implementation at \url{https://lua.org/}),
uses the default libc's allocator as a base for implementing more complex allocation schemes.

We currently do not suffer from this drawback because we use full pre-allocation.

We also have a procedure that allocates many nodes at once, without saving and restoring the ``first free node'' index-pointer each time.
It falls back to the standard approach of allocating one-by-one if there are too few nodes available.

\subsection{Searching for next or previous set bit in mask}

In the implementations of $\underline{\mathrm{next}}$ and $\underline{\mathrm{prev}}$ operations,
we need to find the index of the next/previous bit set in a mask, starting from the specified bit index $i$
(not including $i$) and returning a special ``invalid'' value if there is no such bit.

This can be done via zeroing lower/higher bits (up to, and including, the bit indexed $i$)
and invoking the find-first-set/find-last-set instruction on the result.
So the problem can be reduced to the question of how to efficiently zero lower of higher bits up to the specified index, inclusively.

For zeroing $(i+1)$ lowest bits,
the ``obvious'' solution of shifting right and then left by $(i+1)$ is incorrect:
in C/C++, shifting by bit width is undefined behavior, and non-SIMD
shift instructions of x86-64 ISA treat the shift count modulo $W$ for bit width $W,$
so, for example, shifting a 64-bit value by 64 is equivalent to shifting by $0$
(that is, no-op) instead of zeroing out the value.
We can mask (perform bitwise AND) the value with $((-1) \ll i) \ll 1,$
or, equivalently, with $(-2) \ll i,$
which has one operation less. Above, negative numbers are modulo $2^W,$
where $W$ is the bit width: $(-1)$ denotes all-ones value, $(-2)$ denotes $(W-1)$ leading ones and then one zero bit.

For zeroing higher bits starting with, and including, bit index $i,$
we can use the \textbf{bzhi} instruction from \textbf{BMI2}~\cite{BMI2_a}~\cite{BMI2_b}~\cite{BMI2_c} instruction set extension on x86-64.
If \textbf{bzhi} cannot be used, we can mask (perform bitwise AND) the value with $(1 \ll i) - 1.$

\subsection{Calculating an upper bound on capacity needed for a size and maximum size for a given capacity}

We now want to obtain an upper bound on the maximum number of nodes required for a size $\mathfrak{S}$ of a tree.

Define
\begin{align*}
r = \begin{cases}
    K \Mod C, & K \Mod C \ne 0; \\
    C,        & K \Mod C = 0,
\end{cases}
\end{align*}
the number of bits actually used to discriminate between different children of the root node.

There are two crucial facts:
no level can have more than $\mathfrak{S}$ nodes,
and no level can ``give birth'' to more than $2^C$ nodes ($2^r$ for the root node).

Then, the upper bound on the number of the maximum number of nodes at level $i$
can be calculated as
\begin{align*}
\mathcal{M}_i = \begin{cases}
\min \{ \mathfrak{S}, \, 1 \}, & i = 0; \\
\min \{ \mathfrak{S}, \, 2^r \}, & i = 1; \\
\min \{ \mathfrak{S}, \, 2^C \cdot \mathcal{M}_{i - 1} \}, & i \ge 2.
\end{cases}
\end{align*}

The upper bound on the maximum number of nodes can then be calculated as
$$
\sum\limits_{i=0}^{\mathcal{L} - 1} \mathcal{M}_i.
$$

To perform the inverse computation, that is, to calculate the number of elements
that will definitely fit into a tree with a given capacity, we can use binary
search using the formulae above.

\subsection{Preemption principle and tree restructuring}

We can now calculate the maximum tree size if all index-pointers are 16-bit or 32-bit.
We assign $C=5, \, K=50$ (on MOEX, prices have 14 decimal digits; we allocate an extra higher bit for sign).
The size of node is $48$ bytes for 16-bit index-pointers, $80$ bytes for 32-bit index-pointers.

\begin{tabular}{ ll }
\begin{tabular}{ |c|c| }
 \hline
 \multicolumn{2}{|c|}{ \textbf{16-bit index-pointers} } \\
 \hline\hline
  Tree size & Memory consumption \\
 \hline\hline
  $9 \cdot 10^2$ & 339.05 Kb \\
 \hline
  $9 \cdot 10^3$ & 2.93 Mb \\
 \hline
  $9 \cdot 10^4$ & N/A \\
 \hline
  $9 \cdot 10^5$ & N/A \\
 \hline
\end{tabular}
&
\begin{tabular}{ |c|c| }
 \hline
 \multicolumn{2}{|c|}{ \textbf{32-bit index-pointers} } \\
 \hline\hline
  Tree size & Memory consumption \\
 \hline\hline
  $9 \cdot 10^2$ & 565.08 Kb \\
 \hline
  $9 \cdot 10^3$ & 4.89 Mb \\
 \hline
  $9 \cdot 10^4$ & 43.78 Mb \\
 \hline
  $9 \cdot 10^5$ & 414.57 Mb \\
 \hline
\end{tabular}
\end{tabular}

\medskip

``N/A'' means that the resulting capacity is more than index-pointers of such size are able to address.

It is not unusual for order books to contain more than $10^6$ entries.
We estimate the number of instruments we want to receive market data from to be around 100.
Say we have 8 Gb of memory to spend on the order books.
Note that we are going to need two trees per instrument: one for asks and one for bids.
It is easy to see that the maximum number of entries in a tree that we can afford lies in $[9 \cdot 10^3; \, 9 \cdot 10^4].$

It may seem that we cannot use a trie for managing order books because it needs to contain much more elements than we can afford.
But note that:
\begin{itemize}
    \item in practice, we only need to iterate over no more than 25 best prices;
    \item the situation when the best price goes through more than $9 \cdot 10^3$ non-empty levels (either up or down) in a single trading session is extremely unlikely.
\end{itemize}

So we propose the following solution to handle to handle this:
on $\underline{\mathrm{insert}},$ if the resulting size of the \textit{glass} would be greater that the maximum size,
preempt the new level into a hash table.
For $\underline{\mathrm{min}}$/$\underline{\mathrm{max}}$ and $\underline{\mathrm{next}}$/$\underline{\mathrm{prev}},$
if the result cannot be found in the \textit{glass} (assuming the restrictions above, we can prove that the size of the \textit{glass} in this case is strictly less than the maximum size),
run the costly procedure of flushing the entries from the hash table back to the tree.

More specifically:
\begin{itemize}
    \item define number $\mathcal{S}$ as the maximum size of the \textit{glass} (around $9 \cdot 10^3$ in our case);
    \item define, for a min-\textit{glass} (where the best price is the minimal one),
        \begin{align*}
            \widehat \infty       & = + \infty, \\
            \pi_1 \PriceBetter \pi_2 & = \pi_1 < \pi_2;
        \end{align*}

        for a max-\textit{glass} (where the best price is the maximal one),
        \begin{align*}
            \widehat \infty       & = - \infty, \\
            \pi_1 \PriceBetter \pi_2 & = \pi_1 > \pi_2.
        \end{align*}

        In other words, $\widehat \infty$ means price that is worse than any ``real'' price that a \textit{glass} may contain,
        and $\pi_1 \PriceBetter \pi_2$ means that price $\pi_1$ is better than $\pi_2.$


    \item maintain a hash table that maps ``preempted'' prices to amounts, initially empty;
    \item maintain a number called ``preemption threshold price'', denoted as $\pi_\mathrm{thres}$, initially $\widehat \infty$;

    \item during insertion with price $\pi$:
        \begin{itemize}
            \item[$\star$] if $\pi \PriceBetter \pi_\mathrm{thres},$ insert as usual, except that if the insertion would ``overflow'' the \textit{glass} (the size would be greater than $\mathcal{S}$),
            do not insert it into the \textit{glass}, but instead insert into the hash table and assign $\pi_\mathrm{thres} \longleftarrow \pi.$
            \item[$\star$] otherwise, insert into the hash table;
        \end{itemize}
        We call the operation of inserting into the hash table instead of the \textit{glass} itself a ``preemption''.

    \item during $\underline{\mathrm{find}}$ and $\underline{\mathrm{erase}}$ with price $\pi$:
        if $\pi \PriceBetter \pi_\mathrm{thres},$ perform the operation on the \textit{glass}; otherwise, perform it on the hash table.

    \item as for the $\underline{\mathrm{min}}$/$\underline{\mathrm{max}}$ and $\underline{\mathrm{next}}$/$\underline{\mathrm{prev}}$
    operations: in this setting, we only support $\underline{\mathrm{min}}$ and $\underline{\mathrm{next}}$ for min-\textit{glass},
    and $\underline{\mathrm{max}}$ and $\underline{\mathrm{prev}}$ for max-\textit{glass}.
    Even then, $\underline{\mathrm{next}}$/$\underline{\mathrm{prev}}$ are only supported within best $\mathcal{B}$ prices, $\mathcal{B} < \mathcal{S}.$
    We define the notion of \textit{exceptional situation} as the situation when:
        \begin{itemize}
            \item[$\star$] we need to handle operation $\underline{\mathrm{min}}$/$\underline{\mathrm{max}}$/$\underline{\mathrm{next}}$/$\underline{\mathrm{prev}},$
                and the result of this operation on the \textit{glass} itself (without elements preempted to the hash table) would be $\#,$ and;
            \item[$\star$] $\pi_\mathrm{thres} \ne \widehat \infty$ (or, equivalently, the hash table is not empty).
        \end{itemize}

    In \textit{exceptional situation}, we need to perform a \textit{tree restructure}, which consists of the following:
        \begin{itemize}
            \item[$\star$] calculate the number of entries to un-preempt from the hash table as $n_\mathrm{unpreempt} = \mathcal{S} - \sigma,$
            where $\sigma$ is the current size of the \textit{glass}.
            Note that $n_\mathrm{unpreempt}$ cannot be zero:
            \begin{itemize}
                \item[$\diamond$] for $\underline{\mathrm{min}}$/$\underline{\mathrm{max}}$ operations, \textit{exceptional situation} means $\sigma = 0;$
                \item[$\diamond$] for $\underline{\mathrm{next}}$/$\underline{\mathrm{prev}}$ on price $\pi$, \textit{exceptional situation} means that $\sigma$
                    is the 0-based rank, counting from the best prices to the worst ones, of $\pi$ in the \textit{glass}.
                    As we only support values of $\pi$ within best $\mathcal{B}$ prices, $\sigma \le \mathcal{B} < \mathcal{S};$
            \end{itemize}

            \item[$\star$] select $n_\mathrm{unpreempt}$ elements (or less if the size of the hash table is less)
            with best prices from the hash table. This can be done either via sorting in $O(n \log n),$
            or via ``partial sorting'' in $O(n + k \log k),$ where $n$ is the size of the hash table, $k = n_\mathrm{unpreempt};$

            \item[$\star$] insert those elements to the \textit{glass} and remove them from the hash table;

            \item[$\star$] assign to $\pi_\mathrm{thres}$ the best price remaining in the hash table, or, if the hash table is now empty, $\widehat \infty.$
        \end{itemize}

        We need, then, to perform the operation that caused \textit{exceptional situation} again: now, an \textit{exceptional situation} cannot arise.
\end{itemize}

That's a lot of text, but it boils downs to pretty compact code:

\begin{procedure}[H]
\DontPrintSemicolon
\caption{ob-init()}
\KwData{
    $\pi_\mathrm{thres},$ the preemption threshold price.\;
    $\Gamma,$ the \textit{glass} data structure.\;
    $\chi,$ the hash table.
}
\Begin{
    $\pi_\mathrm{thres} \longleftarrow \widehat{\infty}$\;
    $\underline{\text{glass-init}}(\Gamma)$\;
    $\underline{\text{hash-table-init}}(\chi)$\;
}
\end{procedure}

\begin{procedure}[H]
\DontPrintSemicolon
\caption{ob-insert($\pi, a$)}
\KwData{
    $\pi_\mathrm{thres},$ the preemption threshold price.\;
    $\Gamma,$ the \textit{glass} data structure.\;
    $\chi,$ the hash table.
}
\Begin{
    \uIf{$\pi \PriceBetter \pi_\mathrm{thres}$}{
        \uIf{$\underline{\text{glass-size}}(\Gamma) < \underline{\text{glass-max-size}}(\Gamma)$}{
            $\underline{\text{glass-insert}}(\Gamma, \pi, a)$\;
        }\Else{
            $\underline{\text{hash-table-insert}}(\chi, \pi, a)$\;
            $\pi_\mathrm{thres} \longleftarrow \pi$\;
        }
    }\Else{
        $\underline{\text{hash-table-insert}}(\chi, \pi, a)$\;
    }
}
\end{procedure}

\begin{procedure}[H]
\DontPrintSemicolon
\caption{ob-erase($\pi$)}
\KwData{
    $\pi_\mathrm{thres},$ the preemption threshold price.\;
    $\Gamma,$ the \textit{glass} data structure.\;
    $\chi,$ the hash table.
}
\Begin{
    \uIf{$\pi \PriceBetter \pi_\mathrm{thres}$}{
        $\underline{\text{glass-erase}}(\Gamma, \pi)$\;
    }\Else{
        $\underline{\text{hash-table-erase}}(\chi, \pi)$\;
    }
}
\end{procedure}

\begin{procedure}[H]
\DontPrintSemicolon
\caption{ob-find($\pi$)}
\KwData{
    $\pi_\mathrm{thres},$ the preemption threshold price.\;
    $\Gamma,$ the \textit{glass} data structure.\;
    $\chi,$ the hash table.
}
\Begin{
    \uIf{$\pi \PriceBetter \pi_\mathrm{thres}$}{
        $\underline{\text{glass-find}}(\Gamma, \pi)$\;
    }\Else{
        $\underline{\text{hash-table-find}}(\chi, \pi)$\;
    }
}
\end{procedure}

\begin{procedure}[H]
\DontPrintSemicolon
\caption{ob-best()}
\KwData{
    $\pi_\mathrm{thres},$ the preemption threshold price.\;
    $\Gamma,$ the \textit{glass} data structure.\;
    $\chi,$ the hash table.
}
\Begin{
    \If{$\underline{\text{glass-size}}(\Gamma) = 0$ \And $\pi_\mathrm{thres} \ne \widehat{\infty}$}{
        $\underline{\text{ob-restructure}}()$\;
    }
    \uIf{\textit{this is a min-orderbook}}{
        \Return $\underline{\text{glass-min}}(\Gamma)$\;
    }\Else{
        \Return $\underline{\text{glass-max}}(\Gamma)$\;
    }
}
\end{procedure}

\begin{procedure}[H]
\DontPrintSemicolon
\caption{ob-next-best-after($\pi$)}
\KwData{
    $\pi_\mathrm{thres},$ the preemption threshold price.\;
    $\Gamma,$ the \textit{glass} data structure.\;
    $\chi,$ the hash table.
}
\Begin{
    \uIf{\textit{this is a min-orderbook}}{
        $f \longleftarrow \underline{\text{glass-next}}$\;
    }\Else{
        $f \longleftarrow \underline{\text{glass-prev}}$\;
    }

    $r \longleftarrow f(\Gamma)$\;
    \If{$r = \#$ \And $\pi_\mathrm{thres} \ne \widehat{\infty}$}{
        $\underline{\text{ob-restructure}}()$\;
        $r \longleftarrow f(\Gamma)$\;
    }
    \Return $r$\;
}
\end{procedure}

\begin{procedure}[H]
\DontPrintSemicolon
\caption{ob-restructure()}
\KwData{
    $\pi_\mathrm{thres},$ the preemption threshold price.\;
    $\Gamma,$ the \textit{glass} data structure.\;
    $\chi,$ the hash table.
}
\Begin{
    $\sigma \longleftarrow \underline{\text{glass-size}}(\Gamma)$\;
    $S \longleftarrow \underline{\text{glass-max-size}}(\Gamma)$\;
    \If{$\sigma = S$}{
        \Error{\textit{``ob-next-best-after()'' price too far from best}}\;
    }
    $n_\mathrm{avail} \longleftarrow S - \sigma$\;
    $B \longleftarrow \underline{\text{hash-table-best-n}}(\chi, \min\{ n_\mathrm{avail}, \underline{\text{hash-table-size}}(\chi) \})$\;
    \ForEach{$\langle \pi, a \rangle \in B$}{
        $\underline{\text{glass-insert}}(\chi, \pi, a)$\;
        $\underline{\text{hash-table-erase}}(\chi, \pi)$\;
    }
    \uIf{$\underline{\text{hash-table-size}}(\chi) = 0$}{
        $\pi_\mathrm{thres} \longleftarrow \widehat{\infty}$\;
    }\Else{
        $\langle \pi_0, a_0 \rangle \longleftarrow \underline{\text{hash-table-best-1}}(\chi)$\;
        $\pi_\mathrm{thres} \longleftarrow \pi_0$\;
    }
}
\end{procedure}

\subsection{Exact division of bit offset by $C$}

In internal iterators, we represent the bit offset of a key as a signed number $\kappa$, with depth-$\delta$ iterator having
$$\kappa = C \cdot (\mathcal{L} - 1 - \delta).$$
Thus, an iterator referring to the root node has $\kappa = C \cdot (\mathcal{L} - 1),$ an iterator referring to a pre-leaf node has $\kappa = 0,$
an iterator referring to a post-leaf node has $\kappa = -C.$
This encoding helps us to traverse the tree: we can
\begin{itemize}
    \item adjust the depth up or down by incrementing or decrementing $\kappa$ by $C;$
    \item load the current chunk as $(\underline{\mathrm{key}} \gg \kappa) \, \& \, M,$ where $M = (1 \ll C) - 1$ is a compile-time constant;
    \item insert a chunk $\eta$ into the current position via $\underline{\mathrm{key}}' = \underline{\mathrm{key}} \, | \, (\eta \ll \kappa);$
\end{itemize}
and perform other similar actions.

Unfortunately, during insertion, we need to map $\kappa$ back to the depth $\delta$ in order to calculate the number of nodes to allocate.
If we make $\kappa$ represent offsets in $C$-sized chunks, not in bits, then, in operations involving bit shifts, we would have to shift by $C \cdot \kappa,$ not simply by $\kappa,$
which is much slower.
If we represent offsets via pairs $\langle \kappa, \kappa / C \rangle,$ then we would have to perform arithmetic on two numbers
instead of one when adjusting the depth up or down.

The formula for mapping $\kappa$ back to depth $\delta$ is
$$
\delta = \mathcal{L} - 1 - (\kappa / C).
$$
Note that the division is always exact.

We can use approach from ~\cite{divcnst} to reduce this division to a shift and a multiplication modulo $2^W,$
where $W$ is the bit width of the integer type in which $\kappa$ is represented.
Specifically, we decompose $C$ into a product $$C = 2^\ell \cdot \omega,$$ where $\ell, \omega \in \mathbb{N}$, $\omega$ is odd.
Note that such a decomposition exists and is unique for integer $C > 0$:
set $\ell$ to the exponent of the maximal (integer) power of two that divides $C,$ set $\omega$ to $C / 2^\ell.$

If a division by odd $\omega$ is known to be exact, we can:
\begin{itemize}
    \item in compile-time: calculate $\omega^{-1},$ the inverse element of $\omega$ in $\mathbb{Z}_{2^W};$
    \item in run-time: multiply the dividend by $\omega^{-1}$ modulo $2^W.$
\end{itemize}

Finally, we can calculate the depth as
$$
    \delta = \mathcal{L} - 1 - \left(
        \left(
            \left( \kappa \gg \ell \right)
            \cdot
            \omega^{-1}
        \right)
        \Mod
        2^{W}
    \right).
$$
If $C$ is odd, then $\ell = 0,$ so we do not need the shift.
If $C$ is a power of two, then $\omega = \omega^{-1} = 1,$
so we do not need the multiplication.

\section{Optional features}

\subsection{Cached path}

\subsubsection{Basics}

In this section, bit sequences in \textbf{bold} mean keys, while \underline{underlined} bit sequences refer to nodes corresponding to those sequences.
As a special case, $\underline{\varepsilon}$ refers to the root node.

As we have previously shown, market data exhibits strong sequential locality.
To exploit this, we can cache the path (up to a pre-leaf node) in the trie to the last inserted element.

The cached path consists of:
\begin{itemize}
    \item the last key;
    \item a \textit{path}, which is an array $\rho$ of index-pointers of length $\mathcal{L};$
    \item the number $d \in \mathbb{N}, d \le \mathcal{L}$ representing the actual size of $\rho.$
\end{itemize}
The latter is needed in order to be able to truncate the cached path on $\underline{\mathrm{erase}}$
operation instead of invalidating all of the cached path.

Here is the situation where the cached path is full ($\mathcal{L} = d = 4$):\\
\begin{figure}[H]
\caption{A trie with full cached path}
\centering
\digraph[scale=0.6]{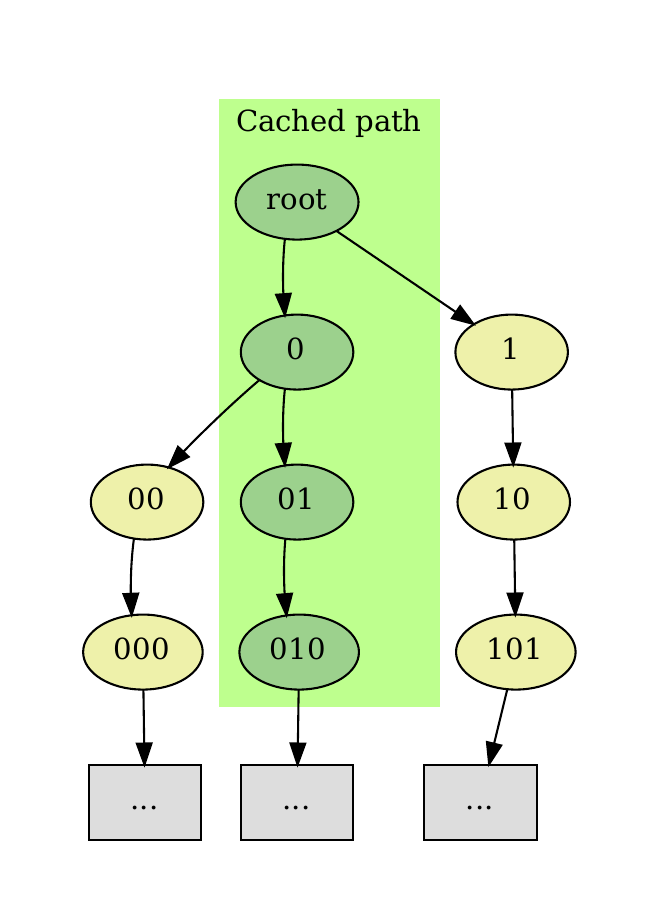}{
graph [ fontname = "DejaVu Math TeX Gyre"; ]
node [ fontname = "DejaVu Math TeX Gyre"; ]
edge [ fontname = "DejaVu Math TeX Gyre"; ]
layout = "dot";
node [
fillcolor = "\MyColor{eef1aa}"; style = "filled";
]
edge[weight=0]
"start" [ label = "root"; fillcolor = "\MyColor{9cd18d}"; ]
"n0" [ label = "0"; fillcolor = "\MyColor{9cd18d}"; ]
"n00" [ label = "00";  ]
"n000" [ label = "000";  ]
"n0000" [ label = "..."; fillcolor = "\MyColor{dddddd}"; style = "filled"; shape = "box"; ]
"n01" [ label = "01"; fillcolor = "\MyColor{9cd18d}"; ]
"n010" [ label = "010"; fillcolor = "\MyColor{9cd18d}"; ]
"n0100" [ label = "..."; fillcolor = "\MyColor{dddddd}"; style = "filled"; shape = "box"; ]
"n1" [ label = "1";  ]
"n10" [ label = "10";  ]
"n101" [ label = "101";  ]
"n1010" [ label = "..."; fillcolor = "\MyColor{dddddd}"; style = "filled"; shape = "box"; ]
n0 -> n00;
n00 -> n000;
n000 -> n0000;
n010 -> n0100;
start -> n1;
n1 -> n10;
n10 -> n101;
n101 -> n1010;
subgraph cluster_cached_path {
label = "Cached path";
color = "\MyColor{beff8e}"; style = "filled";
start -> n0;
n0 -> n01;
n01 -> n010;
}
edge[style=invis,weight=1]
start -> n0 -> {n00, n01} -> {n000, n010};
}
\end{figure}

\subsubsection{Insertion and lookup}

Above, the last operation was insertion of \textbf{010}, so last key is \textbf{010}, $\rho$ is $\langle \underline{\varepsilon}, \underline{0}, \underline{01}, \underline{010} \rangle,$
$d = 3.$ Suppose now we want to insert \textbf{011}:\\
\begin{figure}[H]
\caption{A trie with a to-be-inserted node}
\centering
\digraph[scale=0.6]{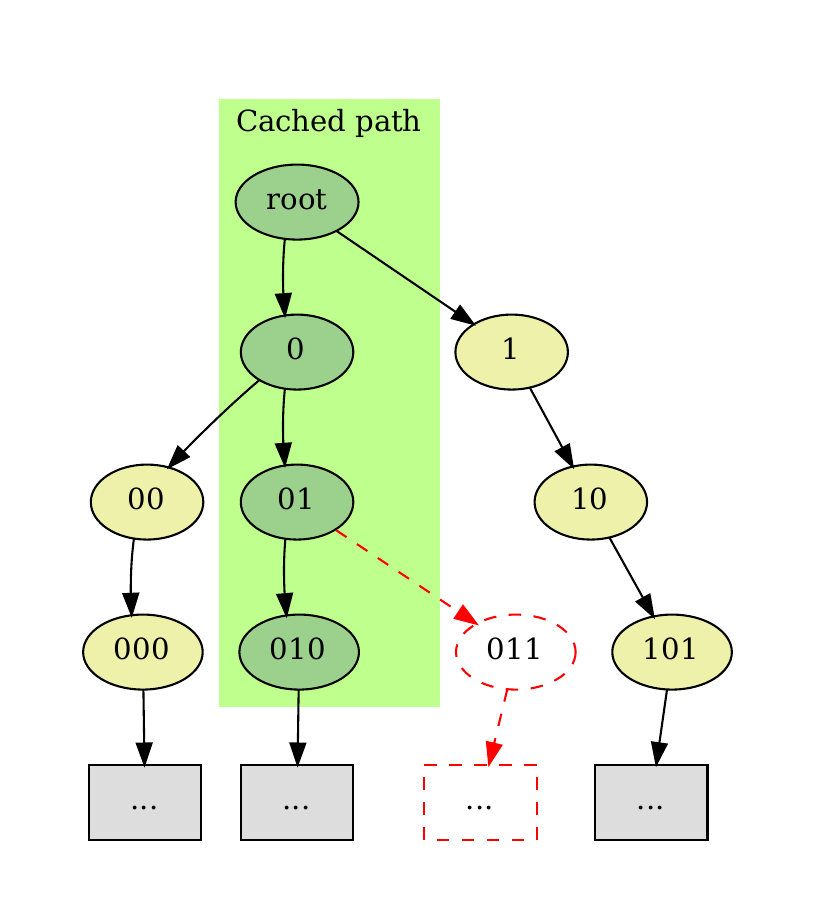}{
graph [ fontname = "DejaVu Math TeX Gyre"; ]
node [ fontname = "DejaVu Math TeX Gyre"; ]
edge [ fontname = "DejaVu Math TeX Gyre"; ]
layout = "dot";
node [
fillcolor = "\MyColor{eef1aa}"; style = "filled";
]
edge[weight=0]
"start" [ label = "root"; fillcolor = "\MyColor{9cd18d}"; ]
"n0" [ label = "0"; fillcolor = "\MyColor{9cd18d}"; ]
"n00" [ label = "00";  ]
"n000" [ label = "000";  ]
"n0000" [ label = "..."; fillcolor = "\MyColor{dddddd}"; style = "filled"; shape = "box"; ]
"n01" [ label = "01"; fillcolor = "\MyColor{9cd18d}"; ]
"n010" [ label = "010"; fillcolor = "\MyColor{9cd18d}"; ]
"n0100" [ label = "..."; fillcolor = "\MyColor{dddddd}"; style = "filled"; shape = "box"; ]
"n011" [ label = "011"; color = "\MyColor{ff0000}"; fillcolor = "\MyColor{ffb0b0}"; style = "dashed"; ]
"n0110" [ label = "..."; color = "\MyColor{ff0000}"; fillcolor = "\MyColor{ffb0b0}"; style = "dashed"; shape = "box"; ]
"n1" [ label = "1";  ]
"n10" [ label = "10";  ]
"n101" [ label = "101";  ]
"n1010" [ label = "..."; fillcolor = "\MyColor{dddddd}"; style = "filled"; shape = "box"; ]
n0 -> n00;
n00 -> n000;
n000 -> n0000;
n010 -> n0100;
n01 -> n011 [ color = "\MyColor{ff0000}"; style = "dashed"; ];
n011 -> n0110 [ color = "\MyColor{ff0000}"; style = "dashed"; ];
start -> n1;
n1 -> n10;
n10 -> n101;
n101 -> n1010;
subgraph cluster_cached_path {
label = "Cached path";
color = "\MyColor{beff8e}"; style = "filled";
start -> n0;
n0 -> n01;
n01 -> n010;
}
edge[style=invis,weight=1]
start -> n0 -> {n00, n01} -> {n000, n010};
}
\end{figure}

The main idea is that we can quickly calculate the number of nodes in the common prefix of the last key and the new key.
Let $k_1, k_2$ be two keys, $W$ bits each. We can calculate the length $\lambda$ of the common prefix, in chunks of $C$ bits, of $k_1$ and $k_2$ as following:
\begin{equation}
    \label{eq:lambdak1k2}
    \lambda = \left\lfloor \frac{\beta - (W - K) + \underline{\mathrm{clz}}_W(k_1 \oplus k_2)}{C} \right\rfloor,
\end{equation}
where bias $\beta = (-K) \Mod C,$ $\oplus$ denotes bitwise XOR operation, and $\underline{\mathrm{clz}}_W(x)$ is the count-leading-zero-bits operation for bit width $W,$
which returns $W$ if $x$ is zero: it counts the number of consecutive zero bits, starting with the most significant one, in $x.$

In the example above, we want to calculate the common prefix of last key \textbf{010} and new key \textbf{011}.
Let us say $W = 8;$ the shape of the tree implies $K = 3, C = 1.$
\begin{itemize}
    \item We calculate $\underline{\mathrm{clz}}_8( \text{\textbf{00000010}} \oplus \text{\textbf{00000011}} ) = 7;$
    \item $\beta = 0,$ $(W - K) = 5,$ so the numerator is $0 - 5 + 7 = 2;$
    \item $\lambda = \lceil 2 / 1 \rceil = 2.$
\end{itemize}
So we jump right into $\rho[\lambda],$ which is $\underline{01};$
it is, indeed, the lowest common ancestor of $\underline{010}$ and $\underline{011}$ (if the latter would be in the tree).

The same sequence of steps can be used to \textit{locate} the key \textbf{011} in the tree if the last key is \textbf{010}.

The count-leading-zeros instruction is readily available on all modern hardware, and is reasonably fast.
Apart from it, the computation compiles down to a XOR, an addition or a subtraction, and division by a constant.
Modern compilers optimize integer division by a constant, using results from ~\cite{divcnst}, down to a sequence of cheaper operations.
For values of $C$ such that $C \le 6$ and $C$ is not a power of two ($C \in \{ 3, 5, 6 \}$),
the sequence only involves single multiplication and single right shift.

We use 6 as a realistic upper bound on $C$ because, on $C > 6,$ the \textit{glass} would occupy an unrealistic amount of memory;
also, the mask would need to contain more than 64 bits, which would slow down common operations on 64-bit hardware.

\subsubsection{Erasure}

Suppose we have a tree (not a trie!), where all leafs are at the same depth, and two paths: the red one goes from the root down to some node (not necessarily a leaf);
the blue one goes from a node (not necessarily the root) down to a leaf. In the picture below, the red path is
$\langle \underline{\varepsilon}, \underline{0}, \underline{01}, \underline{010} \rangle;$
the blue path is $\langle \underline{010}, \underline{0101} \rangle;$
their intersection, the node $\underline{01},$ is colored purple:\\
\begin{figure}[H]
\caption{A tree with red and blue paths}
\centering
\digraph[scale=0.6]{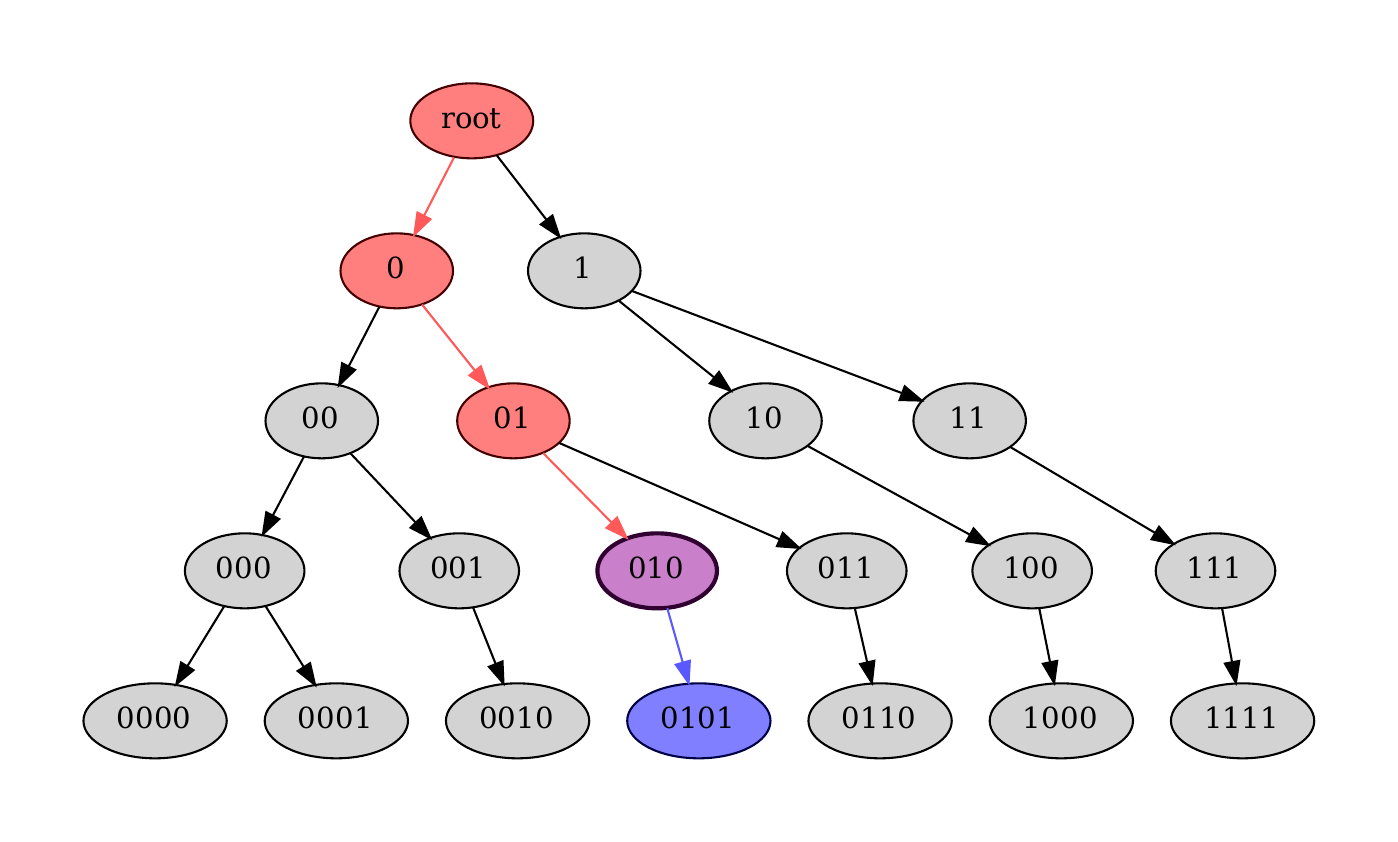}{
graph [ fontname = "DejaVu Math TeX Gyre"; ]
node [ fontname = "DejaVu Math TeX Gyre"; ]
edge [ fontname = "DejaVu Math TeX Gyre"; ]
layout = "dot";
node [
style = "filled";
]
edge[weight=0]
"start" [ label = "root"; color = "\MyColor{440000}"; fillcolor = "\MyColor{ff7f7f}"; ]
"n0" [ label = "0"; color = "\MyColor{440000}"; fillcolor = "\MyColor{ff7f7f}"; ]
"n00" [ label = "00";  ]
"n000" [ label = "000";  ]
"n0000" [ label = "0000";  ]
"n0001" [ label = "0001";  ]
"n001" [ label = "001";  ]
"n0010" [ label = "0010";  ]
"n01" [ label = "01"; color = "\MyColor{440000}"; fillcolor = "\MyColor{ff7f7f}"; ]
"n010" [ label = "010"; color = "\MyColor{300030}"; fillcolor = "\MyColor{c97fc9}"; penwidth = 2; ]
"n0101" [ label = "0101"; color = "\MyColor{000044}"; fillcolor = "\MyColor{7f7fff}"; ]
"n011" [ label = "011";  ]
"n0110" [ label = "0110";  ]
"n1" [ label = "1";  ]
"n10" [ label = "10";  ]
"n100" [ label = "100";  ]
"n1000" [ label = "1000";  ]
"n11" [ label = "11";  ]
"n111" [ label = "111";  ]
"n1111" [ label = "1111";  ]
start -> n0 [ color = "\MyColor{ff5959}"; ];
n0 -> n00;
n00 -> n000;
n000 -> n0000;
n000 -> n0001;
n00 -> n001;
n001 -> n0010;
n0 -> n01 [ color = "\MyColor{ff5959}"; ];
n01 -> n010 [ color = "\MyColor{ff5959}"; ];
n010 -> n0101 [ color = "\MyColor{5959ff}"; ];
n01 -> n011;
n011 -> n0110;
start -> n1;
n1 -> n10;
n10 -> n100;
n100 -> n1000;
n1 -> n11;
n11 -> n111;
n111 -> n1111;
subgraph cluster_cached_path {
label = "Cached path";
color = "\MyColor{beff8e}"; style = "filled";
}
edge[style=invis,weight=1]
;
}
\end{figure}

Suppose we know that:
\begin{itemize}
    \item the distance from a root to a leaf is $\mathbf{L}$ (on the picture above, $\mathbf{L} = 4$);
    \item the length of red path, in edges, is $\mathbf{R}$
        (on the picture above, $\mathbf{R} = 3$);
    \item the length of the blue path, in edges, is $\mathbf{B}$
        (on the picture above, $\mathbf{B} = 1$);
    \item the distance from the root to the lowest common ancestor of the last nodes of red and blue paths, is $\mathbf{Z}$
        (on the picture above, the lowest common ancestor is $\underline{010},$ so $\mathbf{Z} = 3$).
\end{itemize}
The number $\mathbf{I}$ of nodes in the intersection of the red and blue paths can be calculated as follows:
$$
\mathbf{I} = \max \{ 0, \, \mathbf{Z} + \mathbf{B} + 1 - \mathbf{L} \}.
$$
On the picture above, $\mathbf{I} = 1.$
Also note that it turns out we do not even need $\mathbf{R}$ for the calculation of $\mathbf{I}.$

On erasure, our ``red path'' is the cached path, and the ``blue path'' consists of the nodes and edges that have been removed.
We substitute $\mathbf{L} = \mathcal{L},$ and calculate $\mathbf{Z}$ as minimum of:
\begin{itemize}
    \item $d,$ and
    \item the length of the common prefix, in chunks of $C$ bits, of the last key and the erased key (see formula \ref{eq:lambdak1k2}).
\end{itemize}
We then truncate the cached path by $\mathbf{I}$: that is, we assign $$d \longleftarrow \max \{ 0, \, d - \mathbf{I} \}.$$
The $\max$ operator is needed because, if the whole tree has been removed, including the root, $\mathbf{I} = d + 1 > d.$

Let us now briefly go back to our trie examples in the previous subsubsection.
Here is what the cached path looks like after the deletion of the key \textbf{011} (after it has been inserted):\\
\begin{figure}[H]
\caption{A trie after erasure: cached path is truncated}
\centering
\digraph[scale=0.6]{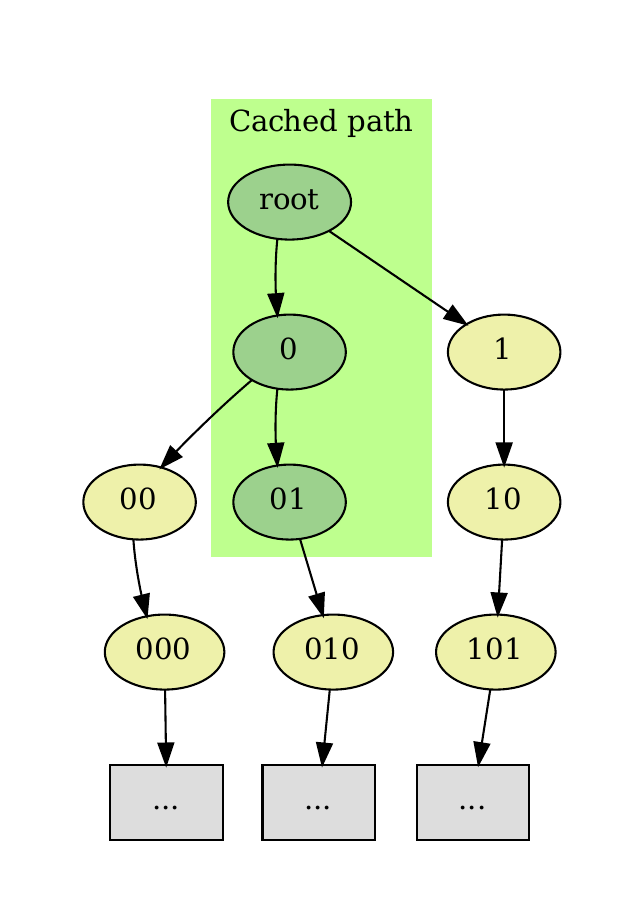}{
graph [ fontname = "DejaVu Math TeX Gyre"; ]
node [ fontname = "DejaVu Math TeX Gyre"; ]
edge [ fontname = "DejaVu Math TeX Gyre"; ]
layout = "dot";
node [
fillcolor = "\MyColor{eef1aa}"; style = "filled";
]
edge[weight=0]
"start" [ label = "root"; fillcolor = "\MyColor{9cd18d}"; ]
"n0" [ label = "0"; fillcolor = "\MyColor{9cd18d}"; ]
"n00" [ label = "00";  ]
"n000" [ label = "000";  ]
"n0000" [ label = "..."; fillcolor = "\MyColor{dddddd}"; style = "filled"; shape = "box"; ]
"n01" [ label = "01"; fillcolor = "\MyColor{9cd18d}"; ]
"n010" [ label = "010";  ]
"n0100" [ label = "..."; fillcolor = "\MyColor{dddddd}"; style = "filled"; shape = "box"; ]
"n1" [ label = "1";  ]
"n10" [ label = "10";  ]
"n101" [ label = "101";  ]
"n1010" [ label = "..."; fillcolor = "\MyColor{dddddd}"; style = "filled"; shape = "box"; ]
n0 -> n00;
n00 -> n000;
n000 -> n0000;
n01 -> n010;
n010 -> n0100;
start -> n1;
n1 -> n10;
n10 -> n101;
n101 -> n1010;
subgraph cluster_cached_path {
label = "Cached path";
color = "\MyColor{beff8e}"; style = "filled";
start -> n0;
n0 -> n01;
}
edge[style=invis,weight=1]
start -> n0 -> {n00, n01} -> n000;
}
\end{figure}

The cached path is now truncated ($d < \mathcal{L}$).

\subsection{Hash table}

In order to further speed up key lookup, the \textit{glass} also supports a hash table (or, rather, a \textit{cache table}).
It maps keys without the last chunk (the least significant $C$ bits) into index-pointers to the pre-leaf the key belongs to.
It uses separate chaining. However, it is different from a standard hash table in the following ways:
\begin{itemize}
    \item A chain is a doubly-linked list, instead of singly-linked, in order to support hard $O(1)$ removal by pointer.
        Insertion is always done into the beginning of a chain;
    \item A lookup only inspects first $J$ elements of a chain, so that lookup is hard $O(1).$
        If a match is found among the first $J$ elements, it returns ``exists'' and the pre-leaf that the key is mapped to.
        If the chain length is less or equal to $J,$ it returns ``doesn't exist''.
        If the chain length is greater then $J,$ it returns ``don't know''.
        $J$ is a compile-time constant which is currently set to $5;$
    \item On a resize (the hash table only supports growing, not shrinking),
        the relative order of the elements within new chains that previously were in the same chain
        is preserved.
        This is done via moving of some elements in the old chains into the beginning of the new chains and
        then reversing the order of elements in the new chains.
        This is done because we believe that recently inserted elements are more likely to be accessed.
\end{itemize}

The next-in-hash-table/previous-in-hash-table index-pointers are embedded into the nodes of the tree, although only used in pre-leaf nodes.
To be able to compare keys during lookup, we also have ``hash table key'' field in every node, although it is also only used in pre-leaf nodes.
The first-in-hash-table index-pointers (we don't need pointers to the end of the chains) are stored in a separate array.
Its size is kept at the largest power of two which is not greater than the tree's capacity.

\subsection{Probability of hash table's ``don't know'' answer}

We can use results from ~\cite{ht_analysis} to calculate the probability of hash table's ``don't know'' answer for a key that is present.

Assume simple uniform hashing.
The probability that a given bucket in a hash table with $n$ elements and $b$ buckets has size $k, 0 \le k \le n,$ is
$$
\mathbf{p}(k) = \binom{n}{k} \left( \frac{1}{b} \right)^k \left( \frac{b-1}{b} \right)^{n-k}.
$$

We can calculate the probability of finding our key in the first $J$ buckets, among $k, 1 \le k \le n$ buckets, as follows:
\begin{align*}
\mathbf{q}(k) = \begin{cases}
    1,   & k \le J; \\
    J/k, & k > J.
\end{cases}
\end{align*}

We can calculate the probability of hash table's ``don't know'' answer, if the key is present, as follows:
$$
p_+ = 1 - \frac{ \sum\limits_{k=1}^{n} \mathbf{p}(k) \mathbf{q}(k) }{ 1 - \mathbf{p}(0) }.
$$
The division is because we only interested in cases where the bucket is not empty (otherwise the key can not be present in this bucket).

The probability of hash table's ``don't know'' answer if the key is \textbf{not} present is just a probability that
a given bucket contains more than $J$ elements:
$$
p_- = \sum\limits_{k=J+1}^{n} \mathbf{p}(k).
$$

We calculated $p_+$ and $p_-$ as functions of $J,$ with
$n = 9210$ (the exact upper bound on the number of elements in a trie with 16-bit index pointer),
$b = 2^{15}$ (the number of buckets in a trie with 16-bit pointers with maximal possible capacity):
\begin{figure}[H]
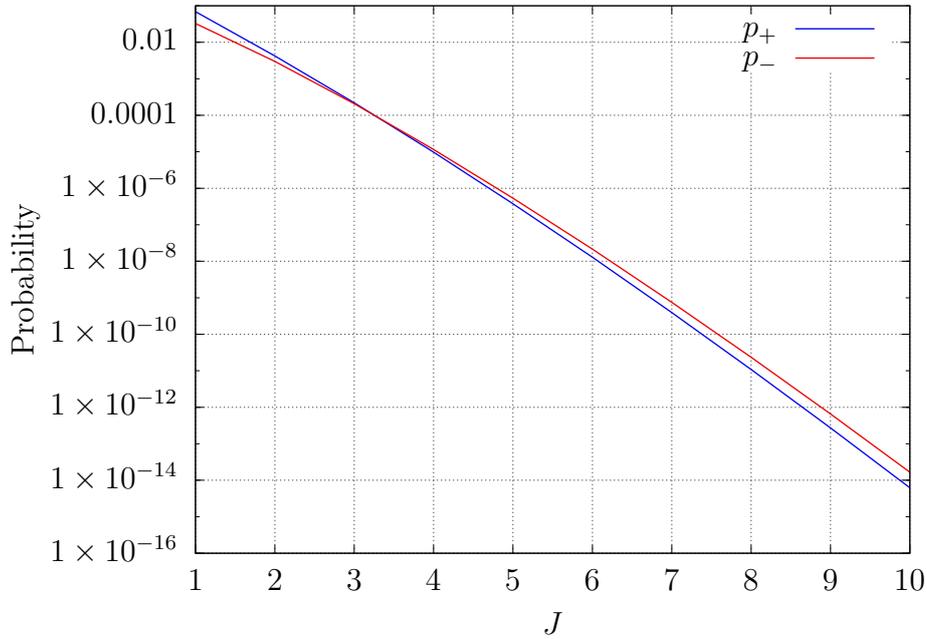

    \caption{Probabilities of hash table's ``don't know'' answer}
    \centering
    \begin{gnuplot}[terminal=epslatex, terminaloptions=color]

        set style line 1 lt 1 lw 2 lc rgb '#0000ee' pt -1
        set style line 2 lt 1 lw 2 lc rgb '#ee0000' pt -1

        set xtics axis
        set ytics axis

        set grid xtics lc rgb '#555555' lw 1 lt 0
        set grid ytics lc rgb '#555555' lw 1 lt 0

        set xlabel "$J$"
        set ylabel "Probability"

        set autoscale xfix

        set logscale y 10

        plot \
            'pdunnoPLUS.txt' using 1:2 with lines ls 1 ti "$p_+$", \
            'pdunnoMINUS.txt' using 1:2 with lines ls 2 ti "$p_-$"

    \end{gnuplot}
\end{figure}

For $J=5,$ $p_+ \approx 3.76 \cdot 10^{-7},$ $p_- \approx 2.14 \cdot 10^{-8}.$

\subsection{Cached iterators to the first and last elements}

\textit{glass} supports caching iterators to the first and last elements.
There are two modes of caching: \textit{eager} and \textit{lazy}.

The iterators can be either valid, point to \textit{end} (meaning the trie is empty), or be in \textit{bad} state (the latter is only possible in the lazy mode).

On insertion, if the inserted element is less/greater than the previous first/last element, the corresponding cached iterator (or both, if the tree was empty) is updated.

Eager and lazy modes differ in the behaviour on erasure (without loss of generality, assume the trie is not empty after
erasure; otherwise, we simply assign \textit{end} to both cached iterators instead):
\begin{itemize}
    \item in eager mode, if the first or last element has been removed,
          the corresponding cached iterator is updated immediately (the relatively costly procedure of finding the first or last element is performed).
          On \underline{first}/\underline{last} query, the corresponding cached iterator is returned;
    \item in lazy mode, if the first or last element has been removed, the corresponding cached iterator is put into \textit{bad} state.
          On \underline{first}/\underline{last} query, if the corresponding cached iterator is not in \textit{bad} state, it is returned;
          otherwise, the procedure of finding the first or last element is performed, the corresponding cached iterator is updated and returned.
\end{itemize}

We can always afford a separate value for iterator's \textit{bad} state because
we always set maximum capacity to at most $2^W - 2,$ where $W$ is the width, in bits, of an index-pointer.

\subsection{Trash encoding}

\textit{Trash encoding} is a way to lower memory usage at a small runtime cost.
\textit{Glass} requires an allocator that produces zeroed out memory at allocation
(in which case the whole new chunk must be zeroed out)
and reallocation that grows an allocated chunk
(in which case the new memory must be zeroed out).
This is required so that we don't need to zero out a new node's mask: it is either freshly-allocated memory, or was zeroed out on erasure.
The default allocator uses \texttt{calloc}/\texttt{realloc}+\texttt{memset}/\texttt{free}.

The idea is that we can use the \texttt{mmap}/\texttt{mremap}/\texttt{munmap} system calls for these operations.
They have granularity of a page (4096 bytes on modern hardware) and, on Linux, they are able to overcommit memory
(produce pages that look zeroed out from the userspace' point of view, but are only associated with physical pages on the first write).
This only works if \texttt{vm.overcommit\_memory} parameter is set to
0 (``heuristic overcommit handling''; it is the default) or
1 (``always overcommit'')
~\cite{linux_overcommit}.

The trash encoding is just a special way to encode an unused node's ``next-free-node'' field:
\begin{itemize}
    \item we introduce a new special value that means ``the next node in the nodes array''; it is encoded as 0;
    \item the special value that means ``no next node'' is encoded as 1;
    \item a reference to the node with index $i$ is encoded as $(i+2).$
        This addition can never overflow because we always set maximum capacity to at most $2^W - 2,$
        where $W$ is the width, in bits, of an index-pointer; and the index of a node is always less than the capacity.
\end{itemize}

When ``next-free-node'' fields are encoded in this way, we do not need to additionally initialize
all the nodes in the beginning and the new nodes after a reallocation,
so the never-used portion of nodes in the end of the nodes array (up to a page boundary) does not consume memory.

The ``next-free-node'' field with value $v$ of an unused node with index $j$ is decoded as follows:
\begin{itemize}
    \item if $v = 0,$ the result is $(j+1);$
    \item it $v = 1,$ the result is ``there's no next node'';
    \item otherwise, the result is $(v-2).$
\end{itemize}

\subsection{Compressed iterators}

If the hash table is used (so that all the pre-leaf nodes contain ``hash table key'' field)
we can only store the $C$ least significant bits of the key in an iterator instead of the full key.
We can then recover the full key via lookup in the node array and a bitwise OR operation, although this is slower than using an ``uncompressed'' iterator.

Assume $C \le 8.$
A compressed iterator would then consist of $\langle i, \, \mathfrak{K} \rangle$ pair, where $i$ is an index-pointer,
$\mathfrak{K}$ is an 8-bit value with the $C$ least significant bits of the key.
For our setup with 16-bit index pointers, $K = 50, \, C = 5,$ this reduces the size of an iterator from 16 bytes to 4 bytes.

\textit{Glass} supports such ``compressed'' iterators (but only if the hash table is used and $C \le 8$).
It provides functions to convert compressed iterator to/from uncompressed ones,
and to get a pointer to the element behind a compressed iterator to perform read/write access through it.

\section{Implementation details}

Our implementation targets C99 with GNU extensions, although it also compiles as C++11 with GNU extensions
in order to be able to compare it against C++' \texttt{std::map}.

We use a custom pre-processor that helps in writing ``X macro''-styled generic code,
and also managing macros (undefining them in the end), including settings that are passed as preprocessor defines.
The code including the glass source must define \texttt{GLASS\_PREFIX};
all functions will be prefixed with it. For example, if \texttt{GLASS\_PREFIX} is \texttt{my\_glass},
the creation function will be called \texttt{my\_glass\_create}.
In order to do this, we define \texttt{GLASS\_NAME(SUFFIX)} macro that concatenates
together (with \texttt{\#\#}) \texttt{GLASS\_PREFIX}, ``\texttt{\_}'' and \texttt{SUFFIX}.
The name of a function then can be written as \texttt{GLASS\_NAME(create)}.
The pre-processor allows us to write \texttt{@create} instead of \texttt{GLASS\_NAME(create)}.
It also serves to reduce error-prone boilerplate related to keeping track of macros that should be
undefined in the end, including the settings definitions.

The preprocessor is called ``ato'', because that's the name of the ``\texttt{@}'' character in Japanese.

\section{Benchmarks}

\subsection{Set and setting}

All measurements were performed on Xiaomi RedmiBook 15 TM2039-44450 laptop.

We have taken the following measures to ensure the benchmarks are as fair as possible:
\begin{itemize}

    \item in order to minimize possible interferences, the benchmark was run in Linux kernel's system console; any other applications, including the X server, were not launched;

    \item \texttt{/sys/devices/system/cpu/cpufreq/policy*/scaling\_governor} policies were set to ``performance'';

    \item before running the benchmark, the driver process has been reniced with ``\texttt{renice -n -20}'';

    \item the driver process was pinned to a single core with ``\texttt{taskset -c 1}'';

    \item timing measurements were taken with ``\texttt{mfence}, \texttt{lfence}, \texttt{rdtsc}, \texttt{lfence}'' sequence of assembly instructions,
          but there is no significant difference in results if \texttt{clock\_gettime(CLOCK\_MONOTONIC, …)} is used instead.

\end{itemize}

In order to be fair, we have also implemented a custom allocator for \texttt{std::map},
which uses the same ``slot allocator'' principle that the \textit{glass}' allocator uses.
It is faster than glibc's allocator used by default because:
\begin{itemize}
    \item slot allocator doesn't do locking;
    \item slot allocator's allocation (assuming no need to allocate a new arena, which is very rare) and deallocation run in hard $O(1)$ time;
    \item glibc's allocator uses red-black trees to manage the list of free chunks, which is slow because of pointer chasing.
\end{itemize}
It is consistently faster in benchmarks.
At first we used a separate slot allocator for each \texttt{std::map} copy, but
the approach with a single common slot allocator turned out not only to be faster,
but also to produce the results that make more sense.

\subsection{Synthetic vs real market data}

For benchmarking, we can use real market data.
Unfortunately, as it is, it does not provide us a way to measure the performance of specific operations
(\underline{insert}, \underline{erase}, \underline{find} of existing/non-existing element, iteration over 25 best prices).

In order to measure these operations separately, we generate synthetic data: to generate a sequence of unique prices,
we use a random number generator to generate differences between successive prices that are distributed just like
the real market data, except that zero difference is prohibited.
We generated synthetic price sequences using the method described above for
\underline{insert}, \underline{erase}, and \underline{find} (of existing/non-existing element)
operations.

\subsection{Amplification}

\textit{Amplification} is an action of replacing an operation in a sequence with multiple identical copies of it.

It only makes sense to amplify read-only operations (\underline{find} and iteration), because
a second \underline{insert} with the same key degenerates to a lookup (which would say the key already exists in the \textit{glass}),
and likewise a second \underline{erase} with the same key degenerates to a lookup (which would say the key is not present in the \textit{glass}).
These modifying operations also modify the cached path, so any time measurements of their amplified copies would not be representative.

Because we can measure the performance of \underline{find} operation with synthetic data,
we only amplify the operation of iteration over the best 25 prices (in the captions of the graphs below referred to as ``iter'').
The amplification is done for the sequence of operations representing the real market data.
We chose to set the amplification coefficient to 100x.

When a read-only operation is amplified, it also makes sense to remove all other read-only operations from the input
in order to be closer to only measuring this operation.
In the context of the previous paragraph, this means that we remove \underline{find} operations from the input
when measuring amplified iteration.

\subsection{Multiple copies}

In order to have a more realistic benchmark, we create multiple copies of the data structure being benchmarked
(either \textit{glass} or \texttt{std::map}) that are operated upon: instead of applying the operation to a
single copy, we apply it to all the copies.

Since in reality we are going to operate upon order books of up to 100 instruments,
benchmarks with multiple copies are more faithful, in particular regarding the behaviors related to CPU caches.

The graphs below are parameterized by the number of copies (from 1 to 32, inclusively).

\subsection{Graphs}

Each test was run with the following number of iterations ($n$ is the number of copies):
\begin{itemize}
    \item $\lfloor \frac{2500}{n} \rfloor$ for tests that use synthetic data;
    \item $\lfloor \frac{7500}{n} \rfloor$ for tests that use the real market data.
\end{itemize}

\begin{figure}[H]
    \caption{Synthetic data: \textbf{insert}}
    \centering
    \begin{gnuplot}[terminal=epslatex, terminaloptions=color]

        set style line 1 lt 1 lw 2 lc rgb '#0000ee' pt -1
        set style line 2 lt 1 lw 2 lc rgb '#ee0000' pt -1

        set xtics axis
        set ytics axis

        set grid xtics lc rgb '#555555' lw 1 lt 0
        set grid ytics lc rgb '#555555' lw 1 lt 0

        set xlabel 'Number of copies'
        set ylabel 'Speedup ratio'

        set autoscale xfix

        set key top left Left reverse

        plot \
            'bench/br_SYNTH_INSERT.txt' using 1:2 with lines ls 1 ti 'Against std::map', \
            'bench/br_SYNTH_INSERT.txt' using 1:3 with lines ls 2 ti 'Against std::map w/ custom allocator'

    \end{gnuplot}
\end{figure}

\begin{figure}[H]
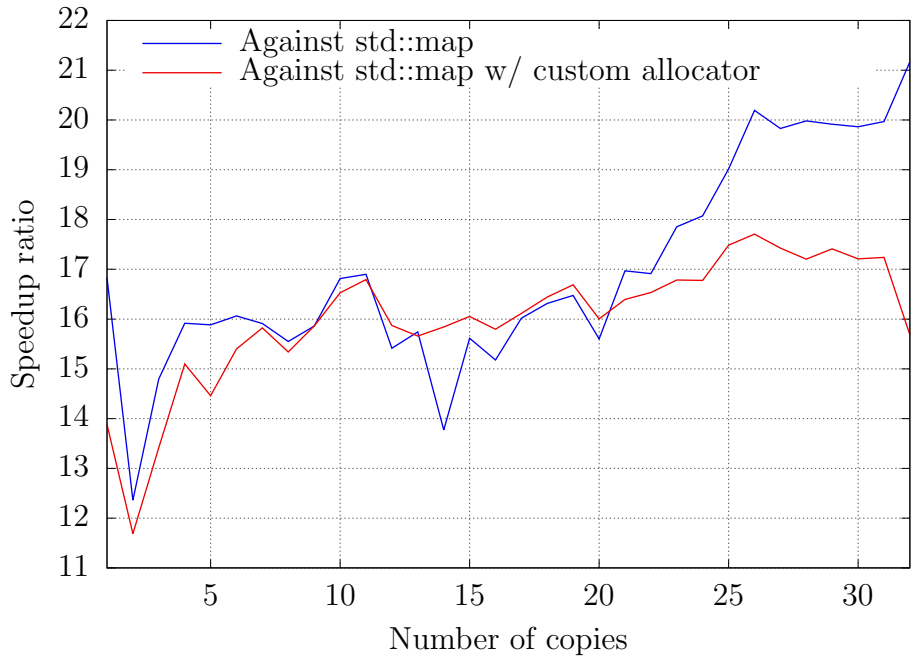

    \caption{Synthetic data: \textbf{erase}}
    \centering
    \begin{gnuplot}[terminal=epslatex, terminaloptions=color]

        set style line 1 lt 1 lw 2 lc rgb '#0000ee' pt -1
        set style line 2 lt 1 lw 2 lc rgb '#ee0000' pt -1

        set xtics axis
        set ytics axis

        set grid xtics lc rgb '#555555' lw 1 lt 0
        set grid ytics lc rgb '#555555' lw 1 lt 0

        set xlabel 'Number of copies'
        set ylabel 'Speedup ratio'

        set autoscale xfix

        set key top left Left reverse

        plot \
            'bench/br_SYNTH_ERASE.txt' using 1:2 with lines ls 1 ti 'Against std::map', \
            'bench/br_SYNTH_ERASE.txt' using 1:3 with lines ls 2 ti 'Against std::map w/ custom allocator'

    \end{gnuplot}
\end{figure}

\begin{figure}[H]
    \caption{Synthetic data: \textbf{find existing}}
    \centering
    \begin{gnuplot}[terminal=epslatex, terminaloptions=color]

        set style line 1 lt 1 lw 2 lc rgb '#0000ee' pt -1
        set style line 2 lt 1 lw 2 lc rgb '#ee0000' pt -1

        set xtics axis
        set ytics axis

        set grid xtics lc rgb '#555555' lw 1 lt 0
        set grid ytics lc rgb '#555555' lw 1 lt 0

        set xlabel 'Number of copies'
        set ylabel 'Speedup ratio'

        set autoscale xfix

        set key top left Left reverse

        plot \
            'bench/br_SYNTH_FIND_E.txt' using 1:2 with lines ls 1 ti 'Against std::map', \
            'bench/br_SYNTH_FIND_E.txt' using 1:3 with lines ls 2 ti 'Against std::map w/ custom allocator'

    \end{gnuplot}
\end{figure}

\begin{figure}[H]
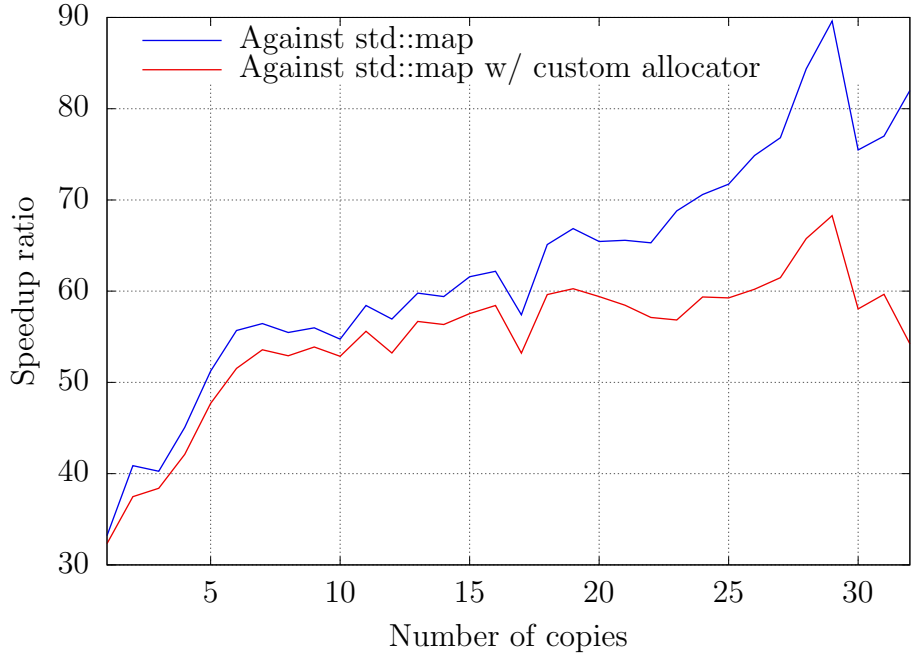

    \caption{Synthetic data: \textbf{find non-existing}}
    \centering
    \begin{gnuplot}[terminal=epslatex, terminaloptions=color]

        set style line 1 lt 1 lw 2 lc rgb '#0000ee' pt -1
        set style line 2 lt 1 lw 2 lc rgb '#ee0000' pt -1

        set xtics axis
        set ytics axis

        set grid xtics lc rgb '#555555' lw 1 lt 0
        set grid ytics lc rgb '#555555' lw 1 lt 0

        set xlabel 'Number of copies'
        set ylabel 'Speedup ratio'

        set autoscale xfix

        set key top left Left reverse

        plot \
            'bench/br_SYNTH_FIND_NE.txt' using 1:2 with lines ls 1 ti 'Against std::map', \
            'bench/br_SYNTH_FIND_NE.txt' using 1:3 with lines ls 2 ti 'Against std::map w/ custom allocator'

    \end{gnuplot}
\end{figure}

\begin{figure}[H]
    \caption{Real market data: \textbf{no amplification}}
    \centering
    \begin{gnuplot}[terminal=epslatex, terminaloptions=color]

        set style line 1 lt 1 lw 2 lc rgb '#0000ee' pt -1
        set style line 2 lt 1 lw 2 lc rgb '#ee0000' pt -1

        set xtics axis
        set ytics axis

        set grid xtics lc rgb '#555555' lw 1 lt 0
        set grid ytics lc rgb '#555555' lw 1 lt 0

        set xlabel 'Number of copies'
        set ylabel 'Speedup ratio'

        set autoscale xfix

        set key top left Left reverse

        plot \
            'bench/br_DATA_all.txt' using 1:2 with lines ls 1 ti 'Against std::map', \
            'bench/br_DATA_all.txt' using 1:3 with lines ls 2 ti 'Against std::map w/ custom allocator'

    \end{gnuplot}
\end{figure}

\begin{figure}[H]
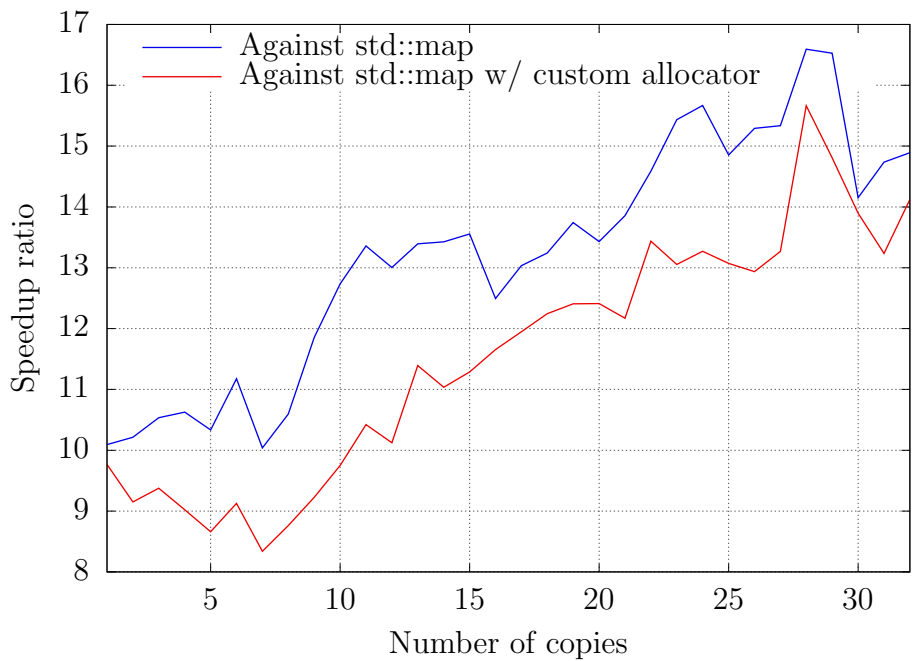

    \caption{Real market data: \textbf{iter amplified 100x}}
    \centering
    \begin{gnuplot}[terminal=epslatex, terminaloptions=color]

        set style line 1 lt 1 lw 2 lc rgb '#0000ee' pt -1
        set style line 2 lt 1 lw 2 lc rgb '#ee0000' pt -1

        set xtics axis
        set ytics axis

        set grid xtics lc rgb '#555555' lw 1 lt 0
        set grid ytics lc rgb '#555555' lw 1 lt 0

        set xlabel 'Number of copies'
        set ylabel 'Speedup ratio'

        set autoscale xfix

        set key top left Left reverse

        plot \
            'bench/br_DATA_iter.txt' using 1:2 with lines ls 1 ti 'Against std::map', \
            'bench/br_DATA_iter.txt' using 1:3 with lines ls 2 ti 'Against std::map w/ custom allocator'

    \end{gnuplot}
\end{figure}

\section{Availability}

The code of our implementation and \LaTeX{} source of this paper
are available at \url{https://github.com/shdown/glass-paper}.
The code is licensed under the MIT license.
The source of this paper is licensed under the Creative Commons BY 4.0 license.

\bibliography{paper}{}

\begin{thebibliography}{10}

\bibitem{linux_overcommit}
{Linux} kernel developers.
\newblock
  \url{https://www.kernel.org/doc/Documentation/vm/overcommit-accounting}.
\newblock [Online; accessed 27-May-2025].

\bibitem{BMI2_c}
AMD.
\newblock {AMD64} technology: {AMD64} architecture programmer’s manual.
  volume 3: General-purpose and system instructions.
\newblock
  \url{https://www.amd.com/content/dam/amd/en/documents/processor-tech-docs/programmer-references/24594.pdf}.
\newblock [Online; accessed 27-May-2025].

\bibitem{pl_rust_map}
Rust documentation.
\newblock {BTreeMap} in std::collections.
\newblock \url{https://doc.rust-lang.org/std/collections/struct.BTreeMap.html}.
\newblock [Online; accessed 27-May-2025].

\bibitem{pl_rust_set}
Rust documentation.
\newblock {BTreeSet} in std::collections.
\newblock \url{https://doc.rust-lang.org/std/collections/struct.BTreeSet.html}.
\newblock [Online; accessed 27-May-2025].

\bibitem{mem_glibc}
Free~Software Foundation.
\newblock Freeing after malloc (the {GNU} {C} library).
\newblock
  \url{https://www.gnu.org/software/libc/manual/html_node/Freeing-after-Malloc.html}.
\newblock [Online; accessed 27-May-2025].

\bibitem{divcnst}
Torbj\"{o}rn Granlund and Peter~L. Montgomery.
\newblock Division by invariant integers using multiplication.
\newblock {\em SIGPLAN Not.}, 29(6):61–72, June 1994.

\bibitem{BMI2_a}
Intel.
\newblock Documentation for \texttt{\_bzhi\_u32} and \texttt{\_bzhi\_u64}
  intrinsics.
\newblock
  \url{https://www.intel.com/content/www/us/en/docs/cpp-compiler/developer-guide-reference/2021-8/bzhi-u32-64.html}.
\newblock [Online; accessed 27-May-2025].

\bibitem{BMI2_b}
Intel.
\newblock Intel® 64 and {IA-32} architectures software developer’s manual,
  volume 2 ({2A}, {2B}, {2C}, \& {2D}): Instruction set reference.
\newblock
  \url{https://cdrdv2-public.intel.com/789581/325383-sdm-vol-2abcd.pdf}.
\newblock [Online; accessed 27-May-2025].

\bibitem{pl_cxx}
{ISO}.
\newblock {ISO\slash IEC JTC1 SC22 WG21 N 4860}: {Programming} languages ---
  {C++}.
\newblock \url{https://isocpp.org/files/papers/N4860.pdf}, 2020.
\newblock [Online; accessed 27-May-2025].

\bibitem{ht_analysis}
R.~Christopher Lacher.
\newblock Hash table analysis (course material).
\newblock \url{https://www.cs.fsu.edu/~lacher/courses/notes/hashanalysis.pdf}.
\newblock [Online; accessed 27-May-2025].

\bibitem{linux_rb}
Rob Landley.
\newblock Red-black trees (rbtree) in {Linux}.
\newblock Linux kernel documentation, file ``rbtree.txt''.
  \url{https://www.kernel.org/doc/Documentation/rbtree.txt}.
\newblock [Online; accessed 21-April-2025].

\bibitem{pl_java_treemap}
Oracle.
\newblock {TreeMap (Java Platform SE 8)}.
\newblock
  \url{https://docs.oracle.com/javase/8/docs/api/java/util/TreeMap.html}.
\newblock [Online; accessed 27-May-2025].

\bibitem{pl_java_treeset}
Oracle.
\newblock {TreeSet (Java Platform SE 8)}.
\newblock
  \url{https://docs.oracle.com/javase/8/docs/api/java/util/TreeSet.html}.
\newblock [Online; accessed 27-May-2025].

\bibitem{mem_lua}
Waldemar~Celes Roberto~Ierusalimschy, Luiz Henrique de~Figueiredo.
\newblock {Lua} 5.3 reference manual --- {lua\_Alloc}.
\newblock \url{https://www.lua.org/manual/5.3/manual.html#lua_Alloc}.
\newblock [Online; accessed 27-May-2025].

\end{thebibliography}
\bibliographystyle{plain}

\end{document}